\documentclass[12pt]{revtex4}
\usepackage{amssymb}
\usepackage{latexsym}
\usepackage{epsfig}
\begin{document}                             %updated 12 June 2011%

\title{ Constructing warm  inflationary model in finite temperature BIon
 }

\author{M. R. Setare $^{1}$\footnote{rezakord@ipm.ir}, A. Sepehri $^{2}$\footnote{alireza.sepehri@uk.ac.ir}, }
\address{$^1$ Department of Science, Campus of Bijar,
University of Kurdistan, Bijar, Iran.\\ $^2$  Faculty of Physics, Shahid Bahonar University, P.O. Box 76175, Kerman, Iran.
}

\begin{abstract}
We study warm inflationary universe model on the BIon in thermal
background. The BIon is a configuration in flat space of a D-brane
and a parallel anti-D-brane connected by a wormhole with F-string
charge. When the  branes and antibranes are well separated and the
brane's spike is far from the antibrane's spike, wormhole isn't
formed however when two branes are close to each other, they can
be connected by a wormhole. In this condition, there exists many
channels for flowing energy from extra dimensions into our
universe and inflation may naturally occur in a warm region. We
present a model that allows all cosmological parameters like the
scale factor $a$, the Hubble parameter $H$ and universe energy density
depend on the shape function and temperature of wormhole in
transverse dimension between two branes. In our model, the
expansion of 4D universe is controlled by the thermal wormhole
between branes and ends up in Big-Rip singularity. We show that at
this singularity, universe would be destroyed and one black
F-string formed. Finally, we test our model against WMAP, Planck
and BICEP2 data and obtain the ripping time. According to
experimental data, $N\simeq 50$ case leads to $n_{s}\simeq 0.96$,
where \emph{N} and $n_{s}$ are the number e-folds and the spectral
index respectively. This standard case may be found in $0.01 <
R_{Tensor-scalar } < 0.3$, where $R_{Tensor-scalar }$ is the
tensor-scalar ratio. At this point, the finite time that Big Rip
singularity occurs is $t_{rip}=29(Gyr)$ for WMAP and Planck data
and $t_{rip}=27.5(Gyr)$ for BICEP2 data. Comparing this time with
the time of Big Rip in brane-antibrane, we find that the wormhole
in BIonic system accelerates the destruction of the universe.

\end{abstract}

 \maketitle
\section{Introduction}
About 20 years ago, Berera has presented a new inflationary model
in which thermal equilibrium is maintained during the inflation
stage.  The result of this warm inflation depends on the presence
of a first order phase transition. In such a case, below $T_{c}$,
there is a temperature barrier which separates the symmetry broken
and unbroken phases. Near $T_{c}$, the region from the top of the
temperature barrier to the symmetry broken minima is somewhat
flat. Furthermore, beyond the inflection point, where the
curvature is positive, thermal fluctuations of the inflaton are
damped. It appears that these two conditions can support a
slow-roll solution for unexceptional values of the coupling
constant, provided that there is also a dissipative component of
sufficient size in the inflaton's equation of motion \cite{m1}.
Until now, many discussions have been done on the warm
inflationary model of universe \cite{m2,m3,m4,m5,m6}. For example,
one paper described the motivation for dissipation and fluctuation
during inflation and then examined their origins from first
principles quantum field theory treatments of inflation models. It
also considered tests for detecting observational signatures of
dissipative processes. In addition it is discussed how particle
physics and cosmology are now working in tandem to push the
boundaries of our knowledge about fundamental physics \cite{m2}.
In another work, the microscopic quantum field theory origins and
the basics of thermal field theory required in the study of warm
inflation dynamics were reviewed. In this research, the
dissipation and stochastic fluctuations were derived of
dissipation coefficients for a variety of quantum field theory
interaction structures relevant to warm inflation, in a form that
can readily be used by model builders \cite{m3}. In another
scenario, some authors have shown that the presence of a thermal
bath warmer than the Hubble scale during inflation decreases the
tensor-to-scalar ratio with respect to the conventional prediction
in supercooled inflation, yielding $r < $8$|n_{t}|$, where $n_{t}$
is the tensor spectral index. Focusing on supersymmetric models at
low temperatures, they have determined consistency relations
between the observables characterizing the spectrum of adiabatic
scalar and tensor modes, both for generic potentials and
particular canonical examples, and which they have compared with
the WMAP and Planck results \cite{m4}. Also, some investigators
have discussed that the presence of even small dissipative effects
at the time when observable scales leave the horizon during
inflation may have a significant effect on the spectrum of
primordial fluctuations in the warm regime, for T and H. This
generically lowers the tensor-to-scalar ratio and yields a
modified consistency relation for warm inflation that may be used
to distinguish it in a model-independent way from the standard
supercooled scenarios if a tensor component is found and
accurately measured \cite{m5}. Some other authors have shown that,
in constructions with additional intersecting D-branes,
brane-antibrane inflation may naturally occur in a warm regime,
such that strong dissipative effects damp the inflaton's motion.
They illustrate this for D3-$\overline{D3}$ inflation in flat
space with additional flavor D7-branes, where for a Coulomb-like
or quadratic hybrid potential a sufficient number of e-folds may
be obtained for perturbative couplings and O(10-$10^{4}$) branes.
This is in clear contrast with the corresponding cold scenarios
\cite{m6}.

 Newly, we considered the origin of warm inflationary model  in the context of brane-antibrane system in Ref.\cite{m7}; however we ignored the role of spikes and wormhole. It might be reasonable to ignore the wormhole in the
ultraviolet where the  branes and antibranes are well separated
and the brane's spike is far from the antibrane's spike, it is
likely that one BIon forms and grows  where the spikes of brane
and antibrane meet each other \cite{m8,m9}. In this condition,
there exists many channels for flowing energy from extra
dimensions into our universe, dominating all other forms of
energy, such that the gravitational repulsion and bringing our
brief epoch of universe to an end in the Big Rip
singularity\cite{m10}. We show that at this singularity, universe
would be destroyed and one black f-string formed. The finite time
that this singularity in BIonic system occurs is much earlier than
the brane-antibrane system.

The outline of the paper is as the following.  In section \ref{o1}, we construct warm inflationary model of universe in BIonic system and show that all cosmological parameters depend on the wormhole's parameters between two branes. Also, in this section, we consider Big Rip singularity and calculate the relation between ripping time and the size of wormhole in extra dimension. In section \ref{o2}, we test our model against the observational data from Planck and WMAP collaborations  and obtain the ripping time. In section \ref{o3}, we consider the signature of finite temperature BIon in BICEP2 results. The last section is devoted to summary and conclusion.

%%%%%%%%%%%%%%%%%%%%%%%%%%%%%%%%%%%%%%%%%%%%%%%%%%%%
\section{ The warm  inflationary model of universe in BIonic system}\label{o1}
In this section, we will consider the role of wormholes in warm
inflationary model of universe and show that they are the main
causes of Big Rip singularity and destruction of universe. To this
end, we will construct four dimensional universe in thermal BIon
and discuss that all cosmological parameters depend on the
temperature of BIon, number of branes and the ripping time.

 To
describe the BIon we specialize to an embedding of the D3-brane
world volume in 10D Minkowski space-time with metric:
\begin{eqnarray}
&& ds^{2} = -dt^{2} + dr^{2} + r^{2}(d\theta^{2} + sin^{2}\theta d\phi^{2}) + \sum_{i=1}^{6}dx_{i}^{2}.
\label{Q1}
\end{eqnarray}
without background fluxes. Choosing the world volume coordinates
of the D3-brane as $\lbrace\sigma^{a}, a=0..3\rbrace$ and defining
$\tau = \sigma^{0},\,\sigma=\sigma^{1}$ the embedding of the
three-brane is given by \cite{m8,m9}:
\begin{eqnarray}
t(\sigma^{a}) = \tau,\,r(\sigma^{a})=\sigma,\,x_{1}(\sigma^{a})=z(\sigma),\,\theta(\sigma^{a})=\sigma^{2},\,\phi(\sigma^{a})=\sigma^{3}
\label{Q2}
\end{eqnarray}
and the remaining coordinates $x_{i=2,..6}$ are constant. There is thus one non-trivial embedding
function $z(\sigma)$ that describes the bending of the brane. Let z be a
transverse coordinate to the branes and $\sigma$ be the radius on the world-volume. The case
of the flat branes is thus $z(\sigma) = 0$.
 The induced metric on the brane is
then:
\begin{eqnarray}
\gamma_{ab}d\sigma^{a}d\sigma^{b} = -d\tau^{2} + (1 + z'(\sigma)^{2})d\sigma^{2} + \sigma^{2}(d\theta^{2} + sin^{2}\theta d\phi^{2})
\label{Q3}
\end{eqnarray}
so that the spatial volume element is $dV_{3}=\sqrt{1 +
z'(\sigma)^{2}}\sigma^{2}d\Omega_{2}$. We impose the two boundary
conditions that $z(\sigma)\rightarrow 0$ for $\sigma\rightarrow
\infty$ and $z'(\sigma)\rightarrow -\infty$ for $\sigma\rightarrow
\sigma_{0}$, where $\sigma_{0}$ is the minimal two-sphere radius
of the configuration. Putting now k units of F-string charge along
the radial direction we obtain \cite{m8,m9}:
\begin{eqnarray}
z(\sigma)= \int_{\sigma}^{\infty} d\acute{\sigma}(\frac{F(\acute{\sigma})^{2}}{F(\sigma_{0})^{2}}-1)^{-\frac{1}{2}}
\label{Q4}
\end{eqnarray}
In finite temperature BIon $F(\sigma)$ is given by
\begin{eqnarray}
F(\sigma) = \sigma^{2}\frac{4cosh^{2}\alpha - 3}{cosh^{4}\alpha}
\label{Q5}
\end{eqnarray}
where $cosh\alpha$ is determined by following function:
\begin{eqnarray}
cosh^{2}\alpha = \frac{3}{2}\frac{cos\frac{\delta}{3} + \sqrt{3}sin\frac{\delta}{3}}{cos\delta}
\label{Q6}
\end{eqnarray}
with the definitions:
\begin{eqnarray}
cos\delta \equiv \overline{T}^{4}\sqrt{1 + \frac{k^{2}}{\sigma^{4}}},\, \overline{T} \equiv (\frac{9\pi^{2}N}{4\sqrt{3}T_{D_{3}}})T, \, \kappa \equiv \frac{k T_{F1}}{4\pi T_{D_{3}}}
\label{Q7}
\end{eqnarray}
In above equation, T is the finite temperature of BIon, N is the number of D3-branes and $T_{D_{3}}$ and $T_{F1}$ are tensions of brane and fundamental strings respectively. Attaching a mirror solution to Eq. (\ref{Q4}), we construct wormhole configuration. The separation distance $\Delta = 2z(\sigma_{0})$ between the N D3-branes
and N anti D3-branes for a given brane-antibrane wormhole configuration defined by the four
parameters N, k, T and $\sigma_{0}$. We have:
\begin{eqnarray}
\Delta = 2z(\sigma_{0})= 2\int_{\sigma_{0}}^{\infty} d\acute{\sigma}(\frac{F(\acute{\sigma})^{2}}{F(\sigma_{0})^{2}}-1)^{-\frac{1}{2}}
\label{Q8}
\end{eqnarray}
In in
the limit of small temperatures, we obtain:
\begin{eqnarray}
\Delta = \frac{2\sqrt{\pi}\Gamma(5/4)}{\Gamma(3/4)}\sigma_{0}(1 + \frac{8}{27}\frac{k^{2}}{\sigma_{0}^{4}}\overline{T}^{8})
\label{Q9}
\end{eqnarray}

  Let us now discuss the warm  inflationary model of universe in thermal BIon. For this, we
need to compute the contribution of the BIonic system to the four-
dimensional universe energy momentum tensor. We now turn to
obtaining the EM tensor for N D3-branes with an electric field on,
corresponding to k units of electric flux, at non-zero
temperature. We can obtain this from a black D3-F1 brane bound
state geometry assuming that we are in the regime of large N and
$g_{s}N$. We get, \cite{m8},
 \begin{eqnarray}
&& T^{00}=\frac{2T_{D3}^{2}}{\pi T^{4}}\frac{F(\sigma)}{\sqrt{F^{2}(\sigma)-F^{2}(\sigma_{0})}}\sigma^{2}\frac{4cosh^{2}\alpha + 1}{cosh^{4}\alpha} \nonumber \\&&
T^{ii}= -\gamma^{ii}\frac{8T_{D3}^{2}}{\pi T^{4}}\frac{F(\sigma)}{\sqrt{F^{2}(\sigma)-F^{2}(\sigma_{0})}}\sigma^{2}\frac{1}{cosh^{2}\alpha},\,i=1,2,3 \nonumber \\&&T^{44}=\frac{2T_{D3}^{2}}{\pi T^{4}}\frac{F(\sigma)}{F(\sigma_{0})}\sigma^{2}\frac{4cosh^{2}\alpha + 1}{cosh^{4}\alpha}
\label{Q10}
\end{eqnarray}
This equation shows that with increasing temperature in BIonic
system, the energy-momentum tensors increase. This is because that
when spikes of  branes and antibranes are well separated, wormhole
isn't formed and there isn't any channel for flowing energy from
extra dimensions to our universe, however when two branes are
close to each other and connected by a wormhole, temperature
achieves to large values.

The conservation law of energy-momentum relates the tensors calculated in brane-antibrane system with ones associated with the four dimensional universe with the following equation:
 \begin{eqnarray}
&& T^{\mu\nu} = \frac{2}{\sqrt{-det g}}\frac{\delta S}{\delta g_{MN}}\frac{\delta g_{MN}}{\delta g_{\mu\nu}} = T_{MN}\frac{\delta g_{MN}}{\delta g_{\mu\nu}},
\label{Q11}
\end{eqnarray}
where $T^{\mu\nu}$ is the energy-momentum tensor of 4D universe in ten dimensional space-time with the metric of the form:
\begin{equation}
ds^{2} = ds^{2}_{Uni,1} + ds^{2}_{Uni,2} + ds^{2}_{transverse},
\label{Q12}
\end{equation}
Here
 \begin{eqnarray}
&& ds^{2}_{Uni1} = ds^{2}_{Uni2} = -dt^{2} + a(t)^{2}(dx^{2} + dy^{2} + dz^{2}),
\label{Q13}
\end{eqnarray}
and
\begin{eqnarray}
&& ds^{2}_{transverse} = dw^{2} + r^{2}d\phi^{2} ,
\label{Q14}
\end{eqnarray}
where $w$ and $\phi$ are the extra space-like coordinates
perpendicular to two universes, $r =\sqrt{x^{2} + y^{2} + z^{2}}$
and scale factor $a(t)$ is assumed to be functions of time only.
In this model, we introduce two four dimensional  universe that
interact with each other and form a binary system. To connect
these universes, we embed one wormhole into transverse space so
that the embedded surface has equation \cite{m11,m12}:
 \begin{eqnarray}
&&  w' = \frac{dw}{dr} = \pm(\frac{r}{b(r)} - 1)^{-\frac{1}{2}}
\label{Q15}
\end{eqnarray}
 where b(r) is the shape function of wormhole. In this condition, the metric of wormhole can be
written as \cite{m11}:
\begin{eqnarray}
&& ds_{wormhole}^{2} = ds_{transverse}^{2} + dr^{2} \nonumber \\&&= dw^{2} + dr^{2} + r^{2}d\phi^{2}\nonumber \\&& = [1 + (\frac{dw}{dr})^{2}]dr^{2} + r^{2}d\phi^{2}  \nonumber \\&& =
\frac{dr^{2}}{1-b(r)/r} + r^{2}d\phi^{2}
\label{Q16}
\end{eqnarray}
This equation helps us to write:
\begin{eqnarray}
&& ds^{2} = ds^{2}_{Uni,1} + ds^{2}_{Uni,2} + ds^{2}_{wormhole} - dr^{2}  \nonumber \\&&= d\overline{s}^{2}_{Uni,1} + d\overline{s}^{2}_{Uni,2} + ds^{2}_{wormhole}  \nonumber \\&& d\overline{s}^{2}_{Uni1} = d\overline{s}^{2}_{Uni2} = -dt^{2} + \tilde{a}(t)^{2}(dx^{2} + dy^{2} + dz^{2})
\label{Q17}
\end{eqnarray}
where we have defined new scale factor $(\tilde{a})^{2} = a^{2} -1/2$. Using the similarity between wormholes in BIonic and universe systems and also, similarity between  Eq. (\ref{Q15}) and Eq. (\ref{Q4}), we can obtain the explicit form of b(r):
 \begin{eqnarray}
&& b = \frac{rG(r_{0})}{G(r)}
\label{Q18}
\end{eqnarray}
where $r_{0}$ is the location of throat and
\begin{eqnarray}
G(r) =  r^{2}\frac{4cosh^{2}\alpha - 3}{cosh^{4}\alpha}
\label{Q19}
\end{eqnarray}
We also can rewrite Eq. (\ref{Q16}) as following:
\begin{eqnarray}
&& ds_{wormhole}^{2} = [1 + (\frac{dr}{dw})^{2}]dw^{2} + r^{2}d\phi^{2}  \nonumber \\&&= c^{2}(r)dw^{2} + r^{2}d\phi^{2}  \nonumber \\&&= [\frac{G^{2}(r_{0})-G^{2}(r)}{G^{2}(r_{0})}]dw^{2} + r^{2}d\phi^{2} \nonumber \\&& = c^2dw^{2} + r^{2}d\phi^{2}
\label{Q20}
\end{eqnarray}

To obtain the energy- momentum tensor in this system, we use of the Einstein's field equation in
presence of fluid flow  that reads as:
\begin{equation}
{R_{ij}} - \frac{1}{2}{{\mathop{\rm g}\nolimits} _{ij}}R = k{T_{ij}}.
\label{Q21}
\end{equation}
Setting the solution of this equation with the line element of Eq. (\ref{Q17}) in the conservation law  of energy-momentum tensor in Eq. (\ref{Q10}), employing Eq. (\ref{Q11}) and assuming $\dot{c} = \frac{dc}{dr}\frac{dr}{dt}$ yield:
\begin{eqnarray}
&& kT_1^1 = \frac{{5\ddot{ \tilde{a}}}}{\tilde{a}} + \frac{{{3{\dot{\tilde{a}}}^2}}}{{{\tilde{a}^2}}} + \frac{{5\dot{ \tilde{a}}\dot
c}}{{\tilde{a}c}} + \frac{{\ddot c}}{c}\nonumber\\
&& =\frac{{5\ddot{ \tilde{a}}}}{\tilde{a}} + \frac{{{3{\dot{\tilde{a}}}^2}}}{{{\tilde{a}^2}}} +\frac{{5\dot{ \tilde{a}}}}{{\tilde{a}}\pi T^{4}}\frac{G(r)}{\sqrt{G^{2}(r)-G^{2}(r_{0})}}r^{2}\dot{r}\frac{1}{cosh^{2}\alpha}\nonumber\\
&& + \frac{1}{\pi T^{4}}\frac{G'(r)G(r_{0})}{G^{2}(r)-G^{2}(r_{0})}r^{2}(\dot{r})^{2}\frac{1}{cosh^{2}\alpha}\nonumber\\
&& = -\gamma^{11}\frac{8T_{D3}^{2}}{\pi T^{4}}\frac{F(\sigma)}{\sqrt{F^{2}(\sigma)-F^{2}(\sigma_{0})}}\sigma^{2}\frac{1}{cosh^{2}\alpha},  \nonumber\\
&& kT_2^2 = \frac{{5\ddot{ \tilde{a}}}}{\tilde{a}} + \frac{{{3{\dot{\tilde{a}}}^2}}}{{{\tilde{a}^2}}} + \frac{{5\dot{ \tilde{a}}\dot
c}}{{\tilde{a}c}} + \frac{{\ddot c}}{c}\nonumber\\
&& =\frac{{5\ddot{ \tilde{a}}}}{\tilde{a}} + \frac{{{3{\dot{\tilde{a}}}^2}}}{{{\tilde{a}^2}}} +\frac{{5\dot{ \tilde{a}}}}{{\tilde{a}}\pi T^{4}}\frac{G(r)}{\sqrt{G^{2}(r)-G^{2}(r_{0})}}r^{2}\dot{r}\frac{1}{cosh^{2}\alpha}\nonumber\\
&& + \frac{1}{\pi T^{4}}\frac{G'(r)G(r_{0})}{G^{2}(r)-G^{2}(r_{0})}r^{2}(\dot{r})^{2}\frac{1}{cosh^{2}\alpha}\nonumber\\
&& =  -\gamma^{22}\frac{8T_{D3}^{2}}{\pi T^{4}}\frac{F(\sigma)}{\sqrt{F^{2}(\sigma)-F^{2}(\sigma_{0})}}\sigma^{2}\frac{1}{cosh^{2}\alpha}, \nonumber\\
&&  kT_3^3 = \frac{{5\ddot{ \tilde{a}}}}{\tilde{a}} + \frac{{{3{\dot{\tilde{a}}}^2}}}{{{\tilde{a}^2}}} + \frac{{5\dot{ \tilde{a}}\dot
c}}{{\tilde{a}c}} + \frac{{\ddot c}}{c} \nonumber\\
&& =\frac{{5\ddot{ \tilde{a}}}}{\tilde{a}} + \frac{{{3{\dot{\tilde{a}}}^2}}}{{{\tilde{a}^2}}} +\frac{{5\dot{ \tilde{a}}}}{{\tilde{a}}\pi T^{4}}\frac{G(r)}{\sqrt{G^{2}(r)-G^{2}(r_{0})}}r^{2}\dot{r}\frac{1}{cosh^{2}\alpha}\nonumber\\
&& + \frac{1}{\pi T^{4}}\frac{G'(r)G(r_{0})}{G^{2}(r)-G^{2}(r_{0})}r^{2}(\dot{r})^{2}\frac{1}{cosh^{2}\alpha}\nonumber\\
&& =  -\gamma^{33}\frac{8T_{D3}^{2}}{\pi T^{4}}\frac{F(\sigma)}{\sqrt{F^{2}(\sigma)-F^{2}(\sigma_{0})}}\sigma^{2}\frac{1}{cosh^{2}\alpha}, \nonumber\\
&&  kT_4^4 =  \frac{{6\ddot {\tilde{a}}}}{\tilde{a}} + \frac{{6{{\dot {\tilde{a}}}^2}}}{{{\tilde{a}^2}}}\nonumber\\&& = \frac{2T_{D3}^{2}}{\pi T^{4}}\frac{F(\sigma)}{F(\sigma_{0})}\sigma^{2}\frac{4cosh^{2}\alpha + 1}{cosh^{4}\alpha}, \nonumber\\
&& kT_{10}^{10} = \frac{{6{{\dot {\tilde{a}}}^2}}}{{{\tilde{a}^2}}} + \frac{{6\dot {\tilde{a}}\dot c}}{{\tilde{a}c}} \nonumber\\&& = \frac{{6{{\dot {\tilde{a}}}^2}}}{{{\tilde{a}^2}}} + \frac{{6\dot{ \tilde{a}}}}{{\tilde{a}}\pi T^{4}G(r_{0)}}\frac{G(r)}{\sqrt{G^{2}(r)-G^{2}(r_{0})}}r^{2}\dot{r}\frac{1}{cosh^{2}\alpha}\nonumber\\
&& =\frac{2T_{D3}^{2}}{\pi T^{4}F(\sigma_{0})}\frac{F(\sigma)}{\sqrt{F^{2}(\sigma)-F^{2}(\sigma_{0})}}\sigma^{2}\frac{4cosh^{2}\alpha + 1}{cosh^{4}\alpha} \label{Q22}
\end{eqnarray}

 The higher-dimensional stress-energy tensor will be assumed to be that of a perfect fluid and of the form:
\begin{equation}
T_i^j = {\mathop{\rm diag}\nolimits} \left[ { - p, - p, - p, - \bar{p}, - p, - p, - p, \rho } \right],
\label{Q23}
\end{equation}
where $\bar{p}$ is the pressure in the extra space-like dimension. In above equation, we are allowing the pressure in the extra
dimension to be different, in general, from the pressure in the 3D space. Hence, this stress-energy tensor describes a
homogeneous, anisotropic perfect fluid in ten dimensions. By adopting the metric ansatzs in  (\ref{Q12}) and (\ref{Q17}), the conservation law in (\ref{Q11}) and (\ref{Q22}) and the perfect fluid stress-energy tensor in (\ref{Q23}), the field equations are of the form:
\begin{eqnarray}
&& -p = \frac{{5\ddot{ \tilde{a}}}}{\tilde{a}} + \frac{{{3{\dot{\tilde{a}}}^2}}}{{{\tilde{a}^2}}} + \frac{{5\dot{ \tilde{a}}\dot
c}}{{\tilde{a}c}} + \frac{{\ddot c}}{c}\nonumber\\
&& =\frac{{5\ddot{ \tilde{a}}}}{\tilde{a}} + \frac{{{3{\dot{\tilde{a}}}^2}}}{{{\tilde{a}^2}}} +\frac{{5\dot{ \tilde{a}}}}{{\tilde{a}}\pi T^{4}}\frac{G(r)}{\sqrt{G^{2}(r)-G^{2}(r_{0})}}r^{2}\dot{r}\frac{1}{cosh^{2}\alpha}\nonumber\\
&& + \frac{1}{\pi T^{4}}\frac{G'(r)G(r_{0})}{G^{2}(r)-G^{2}(r_{0})}r^{2}(\dot{r})^{2}\frac{1}{cosh^{2}\alpha}\nonumber\\
&& = -\gamma^{11}\frac{8T_{D3}^{2}}{\pi T^{4}}\frac{F(\sigma)}{\sqrt{F^{2}(\sigma)-F^{2}(\sigma_{0})}}\sigma^{2}\frac{1}{cosh^{2}\alpha}\label{Q24}\\
&& -\bar{p} =  \frac{{6\ddot {\tilde{a}}}}{\tilde{a}} + \frac{{6{{\dot {\tilde{a}}}^2}}}{{{\tilde{a}^2}}}\nonumber\\&& = -\frac{2T_{D3}^{2}}{\pi T^{4}}\frac{F(\sigma)}{F(\sigma_{0})}\sigma^{2}\frac{4cosh^{2}\alpha + 1}{cosh^{4}\alpha}\label{Q25}\\
&& \rho = \frac{{6{{\dot {\tilde{a}}}^2}}}{{{\tilde{a}^2}}} + \frac{{6\dot {\tilde{a}}\dot c}}{{\tilde{a}c}} \nonumber\\&& = \frac{{6{{\dot {\tilde{a}}}^2}}}{{{\tilde{a}^2}}} + \frac{{6\dot{ \tilde{a}}}}{{\tilde{a}}\pi T^{4}}\frac{G(r)}{\sqrt{G^{2}(r)-G^{2}(r_{0})}}r^{2}\dot{r}\frac{1}{cosh^{2}\alpha}\nonumber\\
&& =\frac{2T_{D3}^{2}}{\pi T^{4}}\frac{F(\sigma)}{\sqrt{F^{2}(\sigma)-F^{2}(\sigma_{0})}}\sigma^{2}\frac{4cosh^{2}\alpha + 1}{cosh^{4}\alpha} \label{Q26}
\end{eqnarray}
where  we set the higher-dimensional
coupling constant equal to one, k = 1. The equations (\ref{Q24}, \ref{Q25}, \ref{Q26}) help us to explain all properties of the current universe in terms of evolutions in BIonic system. These equations constraint the pressure and density in our universe to wormhole  and express that any increase or decrease in these parameters is due to wormhole evolution in extra dimension.

  We are now at a stage that can enter the properties of warm  inflation into  calculations. In this model, the F-string flux plays the role of inflaton in extra dimension. To get the appropriate F-string
flux on the brane we turn on the world-volume vector field
strength component $F_{01}$, where $F_{01}=\frac{\partial
A_{0}}{\partial x_{1}} -\frac{\partial A_{1}}{\partial x_{0}}  $.
Thus, in four dimensional universe, inflaton is a vector field and
can be introduced by $B_{i} = \frac{A_{i}}{a}$, where a is the
scale factor of universe.  One inflationary model of non-minimally
coupled vector fields in an isotropic and homogeneous universe
(FRW) was presented in \cite{m13}. Vector fields in this scenario
behave in the same way as a minimally coupled scalar field. A
triplet of mutually orthogonal vector fields $B_{i}^{a}$
 are introduced
in \cite{m14}. This vector fields are constrained by
orthogonality:
\begin{equation}
\sum B_{i}^{a}B_{i}^{b} = \vert B \vert^{2}\delta_{a}^{b}.
\label{Q27}
\end{equation}
and
\begin{equation}
\sum B_{i}^{a}B_{j}^{b} = \vert B \vert^{2}\delta_{i}^{j}.
\label{Q28}
\end{equation}
Using these fields and following calculations in \cite{m7}, the
dynamic of warm inflation in spatially flat ten dimensional
space-time is presented by this equation:
\begin{eqnarray}
&& \dot{\rho} + 6H( \rho + p) + \bar{H}( \rho + \bar{p})  = -
\Gamma \dot{B}^{2} . \label{Q29}
\end{eqnarray}
Also, with similar calculations we can obtain:
\begin{eqnarray}
\dot{\rho}_{\gamma} + (8H + \bar{H} )\rho_{\lambda}  = \Gamma
\dot{B}^{2}. \label{Q30}
\end{eqnarray}
where $\rho_{\lambda}$
   is energy density of the radiation, $H=\frac{\dot{\overline{a}}}{a}$, $\overline{H}=\frac{\dot{c}}{c}$ and $\Gamma$ is the dissipative coefficient. In the above
equations dot h.h means derivative with respect to cosmic time.
 Relating the 3D and higher-dimensional pressures to the density  (p = $\omega \rho$, $\bar{p}$ = $\bar{\omega} \rho$) and doing some calculations similar to \cite{m7} lead to following equation:

\begin{eqnarray}
&& ( \bar{\omega}^{2} - 2\bar{\omega} - 3\bar{\omega}\omega +4 )H^{2}\dot{H}  \nonumber\\
&& + ( \bar{\omega}^{2} - 6\bar{\omega}\omega +4 )H^{4} \nonumber\\
&& + ( 1+ \bar{\omega} )\dot{H}^{2} - \bar{\omega}H\ddot{H} = -
\Gamma \dot{B}^{2}. \label{Q31}
\end{eqnarray}
Solving  equations (\ref{Q22} and \ref{Q31}) simultaneously  and
performing some algebra for old universe, we find a consistent
solution for the scale factors, vector inflaton and dissipation
coefficient in terms of ripping time:
\begin{eqnarray}
&& r = r_{0} + t_{0} - t \nonumber \\&& \frac{\dot{\tilde{a}}}{\tilde{a}}= \frac{A}{t_{0} - t}\nonumber \\&& A=\frac{2T_{D3}}{\pi G(r_{0})}\frac{(4cosh^{2}\alpha + 1)^{1/2}}{cosh^{2}\alpha}\nonumber \\&& \tilde{a} = (t_{0} - t)^{A} \label{Q32}  \\\nonumber \\&&\sigma^{2} = \sigma_{0}^{2} + t_{0} - t\nonumber \\&& c = \frac{4cosh^{2}\alpha - 3}{G(r_{0})cosh^{4}\alpha}((t_{0} - t)^{2}+2r_{0}(t_{0} - t))^{1/2} \label{Q33} \\\nonumber  \\
&& \Gamma =( \bar{\omega}^{2} - 2\bar{\omega} - 3\bar{\omega}\omega +4 )(\frac{2T_{D3}}{\pi G(r_{0})}\frac{4cosh^{2}\alpha + 1}{cosh^{2}\alpha})^{3}  \nonumber\\
&& + ( \bar{\omega}^{2} - 6\bar{\omega}\omega +4 )(\frac{2T_{D3}}{\pi G(r_{0})}\frac{4cosh^{2}\alpha + 1}{cosh^{2}\alpha})^{4} \nonumber\\
&& + ( 1+ \bar{\omega} )(\frac{2T_{D3}}{\pi G(r_{0})}\frac{4cosh^{2}\alpha + 1}{cosh^{2}\alpha})^{2} - \bar{\omega}(\frac{2T_{D3}}{\pi G(r_{0})}\frac{4cosh^{2}\alpha + 1}{cosh^{2}\alpha})\nonumber\\
&& B = \frac{G^{2}(r_{0})}{G^{2}(r_{0})-G^{2}(r)} \sim
\frac{1}{\Delta}\sim \frac{1}{(t_{0}-t)^{2}}. \label{Q34}
\end{eqnarray}

Equations (\ref{Q32}) and (\ref{Q33}) indicate that both universe
scale factor and wormhole scale factor depend on the finite
temperature of BIon and tension of D3-branes. Naturally, with
increasing temperature,  the energy of system increases and the
effect of interaction between branes can be observed in scale
factors. Also, this equation shows that if A is smaller than zero,
3D scale factor and Hubble parameter at a certain time (t =
$t_{0}$=$t_{rip}$) are infinity. On the other hand, as can be seen
from equation (\ref{Q33}), the value of the scale factor in
transverse direction is zero at $t = t_{0}$ and as a result  Big
Rip singularity won't happen in extra dimension.

The equation (\ref{Q34}) has some interesting results which can be
used to explain the reasons for occurance of Big Rip singularity
in present era of universe. According to these calculations, when
two branes are located at a large distance from each other
($\Delta=\infty$ and t=0), the vector inflaton and wormhole
parameters are almost zero, while approaching the two together,
spikes of two branes meet, wormhole is formed and  the value of
it's parameters increases. In this situation, brane-antibrane
system disappears and consequently one singularity happens in our
four dimensional universe. Another interesting point that comes
out from this equation is that time of this singularity is
proportional to the initial distance between two branes.

Also,  equation (\ref{Q34}) indicates that the dissipative
coefficient increases with temperature of BIon and tends to
infinity at ripping time. Also, dissipative coefficient is
controlled by the wormhole parameters in extra dimension. This
means that the warm inflation can be affected by phenomenological
events in extra dimension and other universe.

Inserting the solutions in (\ref{Q32} and \ref{Q33}) back into
Eq.(\ref{Q30}), we obtain the energy density of radiation in terms
of ripping time:
\begin{equation}
\rho_{\lambda}\sim \frac{\Gamma}{(8A+B)(t-t_{0})^{4}} \label{Q35}
\end{equation}
As is obvious from Eq.(\ref{Q35}), the energy density of
radiation, increases with time and tends to infinity at Big Rip
singularity. This is because that with moving two branes towards
each other, radiation of inflaton from wormhole grows and acquires
the large values near the colliding point. Another interesting
result that can be deduce from this equation is that the radiating
energy density depends on the temperature of BIon and any
enhancement or decrease in this density can be a signature of some
interactions between two universes in extra dimension.

Now, the main question arises that what is the fate of
universe-wormhole system after the Big Rip singularity? To answer
to this question, we consider the matching of finite temperature
BIon and black f-string at corresponding point\cite{m9}. The
supergravity solution for k coincident non-extremal black
F-strings lying along the z direction is:
\begin{eqnarray}
&& ds^{2} = H^{-1}(-f dt^{2} + dz^{2})+ f^{-1}dr^{2} + r^{2}d\Omega_{7}^{2}\nonumber\\
&& e^{2\phi} = H^{-1},\: B_{0} = H^{-1}-1,\nonumber\\
&& H = 1 +
\frac{r_{0}^{6}sinh^{2}\alpha}{r^{6}},\:f=1-\frac{r_{0}^{6}}{r^{6}}
\label{Q36}
\end{eqnarray}
here written in the string frame. From this the mass density along
the z direction can be found using Ref.\cite{m15}:
\begin{eqnarray}
&& \frac{dM_{F1}}{dz} = \frac{3^{5}T_{D3}^{2}(1 + 6cosh^{2}\alpha)}{2^{7}\pi^{3}T^{6}cosh^{6}\alpha}\nonumber\\
&&k^{2} = \frac{3^{12}T_{D3}^{4}(-1 +
cosh^{2}\alpha)}{2^{12}\pi^{6}T_{F1}^{2}T^{12}cosh^{10}\alpha},\:T_{F1}
= \frac{1}{2\pi l_{s}^{2}}\label{Q37}
\end{eqnarray}
For small temperatures one can
expand the mass density as follows:
\begin{eqnarray}
&& \frac{dM_{F1}}{dz} = T_{F1}k +
\frac{16(T_{F1}k\pi)^{3/2}T^{3}}{81T_{D3}}+
\frac{40T_{F1}^{2}k^{2}\pi^{3}T^{6}}{729T_{D3}^{2}}\label{Q38}
\end{eqnarray}
On the other hand, for finite temperature BIon, we have \cite{m9}:
\begin{eqnarray}
&& M_{BIon} = \int dV_{3}T^{00}=\int dV_{3}\frac{2T_{D3}^{2}}{\pi
T^{4}}\frac{F(\sigma)}{\sqrt{F^{2}(\sigma)-
F^{2}(\sigma_{0})}}\sigma^{2}\frac{4cosh^{2}\alpha +
1}{cosh^{4}\alpha}\label{Q39}
\end{eqnarray}
and as a result, for small temperature, we obtain:
\begin{eqnarray}
&& \frac{dM_{BIon}}{dz} = T_{F1}k + \frac{3\pi T_{F1}^{2}k^{2} T^{4}}{32T_{D3}^{2}\sigma_{0}^{2}}+
 \frac{7\pi^{2} T_{F1}^{3}k^{3} T^{8}}{512T_{D3}^{4}\sigma_{0}^{4}}\label{Q40}
\end{eqnarray}
Also, using Eqs. (\ref{Q15}), (\ref{Q18}), (\ref{Q19}) and (\ref{Q22}) and doing some calculations, we can obtain the mass density along the w direction:
\begin{eqnarray}
&& \frac{dM_{universe-wormhole}}{dz} = \frac{dM_{universe-wormhole}}{dw}\frac{dw}{dz} = T_{F1}k +
 \frac{3\pi k^{4} g_{s}^{2} l_{s}^{4} T^{4}}{2r_{0}}+\frac{\pi^{8} k^{6} g_{s}^{3} l_{s}^{6} T^{8}}{2T_{D3}r_{0}^{2}}\label{Q41}
\end{eqnarray}
Comparing the mass densities for BIon and universe-wormhole systems to the mass density for the F-strings, we see that in order for the thermal
D3-F1 configuration at $\sigma = \sigma_{0}$ and universe-wormhole st $r = r_{0}$ to behave like k F-strings,  $\sigma_{0}$ and $r_{0}$
should have the following dependence on the temperature:
\begin{eqnarray}
&& \sigma_{0} =
(\frac{\sqrt{kT_{F1}}}{T_{D3}})^{1/2}\sqrt{T}[C_{0} +
C_{1}\frac{\sqrt{kT_{F1}}}{T_{D3}}T^{3}]\label{Q42}
\end{eqnarray}
and
\begin{eqnarray}
&& r_{0} = (\frac{2\pi k g_{s}l_{s}^{1/2}}{\sqrt{T_{D3}}})T[F_{0} + F_{1}\frac{2\pi k g_{s}l_{s}^{1/2}}{\sqrt{T_{D3}}}T^{3} +
 F_{2}\frac{4\pi^{2} k^{2} g_{s}^{2}l_{s}}{T_{D3}}T^{6} ]\label{Q43}
\end{eqnarray}
where $T_{F1} = 4k\pi^{2}T_{D3}g_{s}l_{s}^{2}$, $C_{0}$, $C_{1}$,
$F_{0}$, $F_{1}$ and $F_{2}$ are numerical coefficients which can
be determined by requiring that the $T^{3}$ and $T^{6}$ terms in
Eqs. (\ref{Q38}), (\ref{Q40}) and (\ref{Q41}) agree. These
calculations show that when the wormhole grows, there exist more
channels for flowing energy from extra dimension to other four
dimensions. Increasing this energy leads to the destruction of
universe and the formation of the black f-string.

\section{Considering the signature of finite temperature BIon in Planck and WMAP9 data}\label{o2}
In previous section, we propose a model that allows to consider the warm inflation of universe in BIonic system. Here, we examine the correctness of our model
 with observation data and obtain some important parameters like ripping time. Using (\ref{Q32}), the number of e-folds\cite{m16,m17} may be found:
\begin{eqnarray}
&&N=\int_{t_{B}}^{t} H dt=
|A|[ln\frac{t_{rip}-t_{B}}{t_{rip}-t}]\nonumber \\&&
=\frac{2T_{D3}}{\pi G(r_{0})}\frac{(4cosh^{2}\alpha +
1)^{1/2}}{cosh^{2}\alpha}[ln\frac{t_{rip}-t_{B}}{t_{rip}-t}]
\label{Q44}
\end{eqnarray}
where $t_{B}$ denotes the beginning time of inflation epoch.

In Fig.1, we present the number of e-folds \emph{N} for warm
scenario as a function of the t where t is the age of universe. In this plot, we choose A=-80, , $T = 2^{0}k$, $t_{B}=0$ and $t_{rip}=29(Gyr)$.
 We find that \emph{N}=50 leads to $t_{universe}= 13.5(Gyr)$. This result is compatible with with both Planck and WMAP9 data \cite{m18,m19,m20,m21}.
 It is clear that the number of e-folds \emph{N} is much larger for older universe. This is because, as the age of universe increases, the distance
  between spikes of branes becomes smaller, wormhole is formed and more energy enters into the universe.

\begin{figure}\epsfysize=10cm
{ \epsfbox{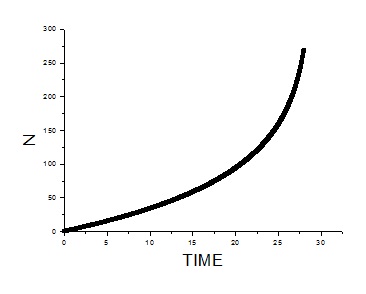}}\caption{The number of e-folds \emph{N} in warm inflation
scenario as a function of $t$ for  A=-80, $T = 2^{0}k$, $t_{B}=0$ and $t_{rip}=29(Gyr)$. } \label{1}
\end{figure}

Another parameters that help us to test our model with
experimental data are the power-spectrum of scalar and tensor
perturbations which presented by \cite{m16,m17}:
\begin{eqnarray}
&&\Delta_{R}^{2}=- (\frac{\Gamma^{3}T^{2}}{36(4\pi)^{3}})^{\frac{1}{2}}\frac{H^{\frac{3}{2}}}{\dot{H}} \nonumber\\
&&\Delta_{T}^{2}= \frac{2H^{2}}{\pi^{2}}, \label{Q45}
\end{eqnarray}
 Using these parameters, we can define the tensor-scalar ratio as \cite{m16,m17}:
\begin{eqnarray}
&& R = -\frac{\Delta_{T}^{2}}{\Delta_{R}^{2}} =
-(\frac{144(4\pi)^{3}}{\Gamma^{3}\pi^{4}
T^{2}})^{\frac{1}{2}}\dot{H}H^{\frac{1}{2}}, \label{Q46}
\end{eqnarray}
In Fig.2, we show the tensor-scalar ratio \emph{R} for warm
scenario as a function of the age of universe. In this plot, we choose A=-80, $t_{B}=0$, $T = 2^{0}k$, c=.01 and $t_{rip}=29(Gyr)$.
 We find that \emph{R}=0.01586 leads to $t_{universe}= 13.5(Gyr)$. This result is compatible with  both Planck and WMAP9 data \cite{m18,m19,m20,m21}.
  Obviously, the tensor-scalar ratio \emph{R} is much larger for older universe. The reason for this is that when two old brane universes  approach  together,
   the size of wormhole increases and it's effect can be observed in observational data.

\begin{figure}\epsfysize=10cm
{ \epsfbox{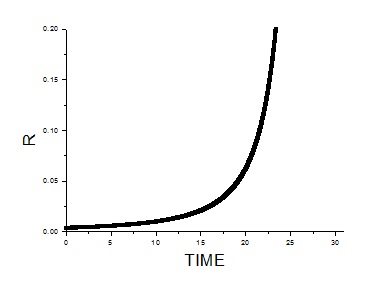}}\caption{The tensor-scalar ratio \emph{R} in warm inflation
scenario as a function of $t$ for A=-80, $T = 2^{0}k$, $t_{B}=0$, C=.01 and $t_{rip}=29(Gyr)$. } \label{2}
\end{figure}

Finally, we compare our model with the Scalar spectral index which
is defined by \cite{m16,m17}:
\begin{equation}
n_{s}-1=-\frac{d ln\Delta_{R}^{2}}{d ln k}=\frac{3}{2}\varepsilon
- (\frac{r'}{4r})\eta \label{Q47}
\end{equation}
Here  k is co-moving wavenumber, $r=\frac{\Gamma}{3H}$ and $\varepsilon$ and $\eta$ are slow-roll parameters of the warm  inflation which are given by:
\begin{eqnarray}
&& \varepsilon =- \frac{1}{H}\frac{d ln H}{dt} \nonumber\\
&& \eta = -\frac{\ddot{H}}{H\dot{H}}, \label{Q48}
\end{eqnarray}
In Fig.3, we show the Scalar spectral index $\emph{n}_{s}$ for
warm scenario as a function of the  age of universe. In this plot,
we choose A=-80 and $t_{B}=0$ and $T = 2^{0}k$. Comparing this
figure with figures(2,3), we find that $N\simeq 50$ case leads to
$n_{s}\simeq 0.96$. This standard case is found in $0.01 <
R_{Tensor-scalar } < 0.22$, which is consistent with both Planck
and WMAP9 data \cite{m18,m19,m20,m21}. At this point, the finite
time that Big Rip singularity occurs is $t_{rip}=29(Gyr)$.
\begin{figure}\epsfysize=10cm
{ \epsfbox{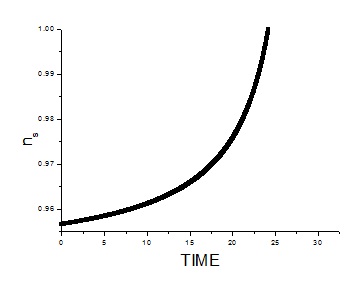}}\caption{The Scalar spectral index $\emph{n}_{s}$ in warm inflation
scenario as a function of $t$ for A=-80, $T = 2^{0}k$, $t_{B}=0$ and $t_{rip}=29(Gyr)$. } \label{3}
\end{figure}
\section{Considering the signature of finite temperature BIon in BICEP2 results}\label{o3}
The BICEP2  collaboration \cite{m22} has announced recently the
detection of primordial B-mode polarization in the cosmic
microwave background (CMB), claiming an indirect observation of
gravitational waves. Such results from
 BICEP2 offer a rare opportunity to directly test
theoretical models, including inflation. The tensor fluctuations
in the cosmic microwave background (CMB) temperatures at large
angular scales are larger than those predicted for inflationary
models based on Einstein gravity. Specifically, the ratio of
tensor-to-scalar perturbations reported by BICEP2 collaboration,
$R = 0.2^{+0.07}_{-0.05}$ (or $R = 0.16^{+0.06}_{-0.05}$ after
subtracting an estimated foreground), is larger than the bounds $R
< 0.13$ and $R < 0.11$ reported by Planck\cite{m18,m19} and WMAP9
data \cite{m20,m21}. Fortunately, in our model, all three data
sets are justified. This is because that tensor fluctuations
depend on the finite temperature and number of branes in BIonic
system and can include both smaller quantities of 0.1 and larger
quantities of 0.19. From (\ref{Q7}), it is seen that for small
temperatures, $\delta$ is close to $\frac{\pi}{2}$, so that
\begin{eqnarray}
 cosh^{2}\alpha = \frac{3\sqrt{3}}{2cos\delta} - \frac{1}{2} - \frac{\sqrt{3}}{12}cos\delta - \frac{2}{27} cos^{2}\delta \label{Q49}
\end{eqnarray}
This gives for the function A defined in (\ref{Q32}) the expansion
\begin{eqnarray}
&& A= \frac{4}{849}cos^{4}\delta - \frac{2\sqrt{3}}{324}cos^{3}\delta + (\frac{29}{432})cos^{2}\delta -(\frac{6\sqrt{3}}{54}) cos\delta - \frac{3\sqrt{3}}{4cos\delta} +\frac{27}{4cos^{2}\delta}\nonumber\\
&& \simeq  z_{1}(NT)^{16}+z_{2}(NT)^{12}+z_{3}(NT)^{8}+\nonumber\\
&&z_{4}(NT)^{4}+z_{5}(NT)^{-4}+ z_{6}(NT)^{-8}\label{Q50}
\end{eqnarray}
where $z_{i}$ are constant coefficients, N is the number of branes and T is the temperature. Also, we can expand the dissipative coefficient in terms of A:
\begin{eqnarray}
&& \Gamma = g_{1} A^{4} + g_{2}A^{3} + g_{3} A^{2}
+g_{4}A\label{Q51}
\end{eqnarray}
 Using these parameters, we can write the tensor-scalar ratio in terms of NT :
\begin{eqnarray}
&& R \sim \frac{1}{( d_{1}A^{4.5} + d_{2} A^{3} +d_{3}A^{3/2})T}(\frac{1}{t-t_{rip}})^{5/2}
\nonumber\\
&&\simeq(p_{1}(NT)^{35}+ p_{2}(NT)^{25} +
..)(\frac{1}{t-t_{rip}})^{5/2} \label{Q52}
\end{eqnarray}
 where $p_{i}$ are constant coefficients. Equation (\ref{Q52}) shows that any increase or
 decrease in the number of branes and temperature can cause a significant change in the tensor-scalar ratio.
 For example, if temperature or number of branes is $10^{\frac{1}{70}}$ higher than the value estimated in previous section,
  the ratio of tensor-to-scalar perturbations would be $R \sim 0.20$ which is in agreement with the value measured by BICEP2 collaboration.

 In Fig.4, we show that the tensor-scalar ratio \emph{R} for warm
scenario as a function of the age of universe.
In this plot, we choose A=-35, $t_{B}=0$, $T = 2^{0}k$, $N \sim 10^{2+\frac{1}{70}}$, c=.01 and $t_{rip}=27.5(Gyr)$.
 We find that \emph{R}=0.20 leads to $t_{universe}= 13.5(Gyr)$. This ratio is compatible with  BICEP2  data \cite{m22}.
 Thus our result are consistent with all experimental data from WMAP, Planck and BICEP2 collaborations and thus our model works.

\begin{figure}\epsfysize=10cm
{ \epsfbox{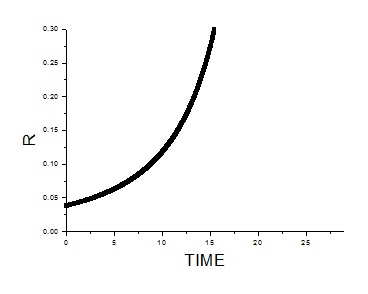}}\caption{The tensor-scalar ratio \emph{R} in warm inflation
scenario as a function of $t$ for A=-35, $T = 2^{0}k$, $N \sim 10^{2+\frac{1}{70}}$, $t_{B}=0$, C=.01 and $t_{rip}=27.5(Gyr)$. } \label{2}
\end{figure}

\section{Summary and Discussion} \label{sum}
In this research, we construct warm  inflation in thermal BIon and show that the energy density, slow-roll, Number of e-fold and
perturbation parameters can be given  in terms of the shape function and the location of the  throat of the wormhole that connects brane and antibrane
in BIonic system. When the  branes and antibranes are well separated and the brane's spike is far from the antibrane's spike, the role of wormhole is ignorable,
 however when  the spikes of brane and antibrane
meet each other, one wormhole would be formed.  According to our results, when the wormhole parameters increase,
 the Number of e-fold, the spectral index and the tensor-scalar ratio  increases and tends to infinity at Big Rip singularity .
  This is because that as the wormhole grows, the effect of interaction between branes on the 4D universe expansion becomes systematically more effective,
   because at larger wormholes, there exist more channels for flowing energy from extra dimension to other four dimensions. Increasing this energy leads to
    the destruction of universe and the formation of the black f-string. We find that $N\simeq 50$ case leads to $n_{s}\simeq 0.96$.
     This standard case may be found in $0.01 < R_{Tensor-scalar } < 0.3$, which agrees with observational data \cite{m18,m19,m20,m21}.
     At this point, the finite time that Big Rip singularity occurs is $t_{rip}=29(Gyr)$ for WMAP and Planck data and $t_{rip}=27.5(Gyr)$ for BICEP2 data.
      This time is much earlier of ripping time in brane-antibrane system in Ref \cite{m7}.

 \end{document}